\documentclass[aps,prl,showpacs,groupedaddress,superscriptaddress,twocolumn]{revtex4}
\usepackage{amsmath}
\usepackage{mathrsfs}
\usepackage{amssymb}
\usepackage[dvips]{graphics,color}
\usepackage{graphicx}
\usepackage{dcolumn}
\usepackage{bm}
\begin{document}
\title{Quantum Noise in the Collective Abstraction Reaction A$+$B$_2\rightarrow$AB$+$B}
\author{H. Jing$^{1,3}$, J. Cheng$^2$, and P. Meystre}
\affiliation{B2 Institute and Department of Physics, The
University of Arizona,
Tucson, Arizona 85721, USA\\
$^2$School of Physical Science and Technology, South China
University of Technology, Guangzhou 510640,
 People's Republic of China\\
 $^3$Department of Physics, Henan Normal University, Xinxiang
453007, People's Republic of China}
\date{\today}
\begin{abstract}
We demonstrate theoretically that the collective abstraction
reaction A+B$_2 \rightarrow$ AB+B can be realized efficiently with
degenerate bosonic or fermionic matter waves. We show that this is
dominated by quantum fluctuations, which are critical in
triggering its initial stages with the appearance of macroscopic
non-classical correlations of the atomic and molecular fields as a
result. This study opens up a promising new regime of quantum
degenerate matter-wave chemistry.
\end{abstract}
\maketitle

The making and probing of ultracold molecular gases has attracted
much attention in recent years \cite{krems}. Starting from an
atomic Bose-Einstein condensate, magnetic Feshbach resonances
and/or optical photoassociation \cite{Paul} can be exploited to
create not only diatomic molecules but also more complex
molecules, such as evidenced by the recent observations of
transient short-lifetime trimers Cs$_3$ \cite{trimer} or tetramers
Cs$_4$ (actually the resonances in inelastic processes)
\cite{tetramer}. In another development of relevance for the
present study, atom-dimer dark states were produced through
coherent photoassociation \cite{winkler}, a process sometimes
called superchemistry or quantum degenerate chemistry
\cite{Heinzen}.

So far, matter-wave superchemistry has concentrated largely on the
coherent combination or decomposition reactions \cite{PD} between
atoms and homonuclear or heteronuclear \cite{Bongs} molecules. In
this Letter we extend these ideas to another type of elementary
chemical reaction, the coherent abstraction reaction (or
bimolecular reactive scattering) A$+$B$_2$$\rightarrow$AB$+$B.
This reaction is an important benchmark system extensively studied
for many years in chemical physics, a particularly noteworthy
contribution being the study by Shapiro and Brumer of the coherent
control of single-molecular photoassociation or bimolecular
collisions via interference of reactive pathways \cite{disp}. One
main result of this Letter is to demonstrate that the coherent
abstraction reaction can be realized and controlled efficiently in
degenerate matter waves by exploiting an atom-molecule dark state.
An important characteristic of this process is that it is
triggered by quantum noise, leading to large shot-to-shot
fluctuations that dominate the initial stages of the reaction.
From a theoretical point of view, this  implies that the
mean-field Gross-Pitaevskii equation is not appropriate to
describe the early stages of the coherent bimolecular reaction,
and can be used only once the product reactant populations become
macroscopic. This is in contrast with single-molecular combination
reactions such as the atom-dimer \cite{Paul,Heinzen} or
atom-trimer conversion \cite{Jing}, and is reminiscent of quantum
or atom optics situations such as the laser, superradiance and
matter-wave superradiance, see for instance Refs. \cite
{QO,Uys,amplifier}.

The basic idea in realizing the collective reaction A+ B$_2
\rightarrow $AB+B in quantum-degenerate gases is to first create
highly excited trimers AB$_2$ via an entrance-channel atom-dimer
Feshbach resonance, and to then photo-dissociate them into a
closed-channel bound dimer and an atom. The use of a dynamical
two-photon resonance scheme involving an intermediate trimer state
permits one to exploit the existence of a coherent population
trapping state (CPT) that prevents the trimer population from
becoming significant throughout the conversion process. Such a
generalized atom-molecule dark state does $not$ exist in other
schemes that involve e.g. an intermediate two-species atomic
state. Note also that this scheme is different from a purely
collision-induced reaction \cite{3body} and from the nondegenerate
single-pair dynamics of reactive scattering \cite{disp}. To our
knowledge this is the first proposal for the quantum control of
matter-wave abstraction reactions, and as such it represents a
promising new step in developing the field of degenerate chemistry
\cite{tetramer,Heinzen}.

\begin{figure}[ht]
\includegraphics[width=8cm]{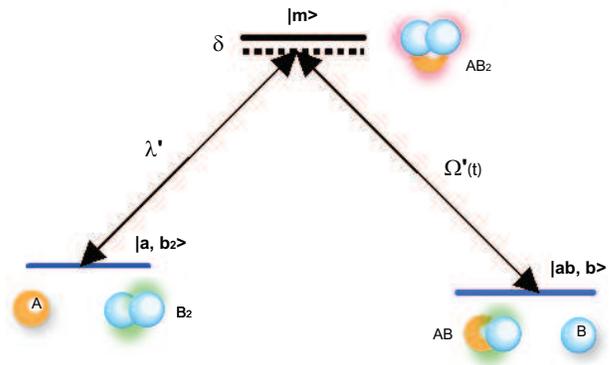}
\caption{(Color online) Schematic illustration of the coherent
abstraction reaction A$+$B$_2\rightarrow$AB$+$B with degenerate
matter waves. Here the atoms A and B can be bosonic or fermionic.}
\end{figure}

Our model system is illustrated by Fig.~1. Denoting the strength
of the A$+$B$_2$$\rightarrow$AB$_2$ coupling with detuning
$\delta$ by $\lambda'_1$, the Rabi frequency of the dissociating
laser by $\Omega'_1$ and its detuning by $\Delta$, the dynamics of
the system is described at the simplest level by the model
Hamiltonian
\begin{eqnarray}
{\cal H}&=&-\int dr \Big \{
 \sum_{i,j}\chi_{i,j}'{\hat \psi}_i^\dag(r){\hat \psi}_j^\dag(r)
 {\hat \psi}_j(r){\hat \psi}_i(r)
 +\delta {\hat \psi}_t^\dag(r) {\hat \psi}_t(r) \nonumber\\
 &+&\lambda'\left [{\hat \psi}
 _t^\dag(r) {\hat \psi}_a(r){\hat \psi}_{b_2}(r)
 +{\rm h.c.}\right ]+(\Delta+\delta){\hat \psi}_{ab}^\dag(r){\hat\psi}_{ab}(r)\nonumber\\
 &-&\left .\Omega'\left [\hat{\psi}_{ab}^\dag(r){\hat \psi}_b^{\dag}(r){\hat
 \psi}_t(r)
 +{\rm h.c.}\right ]
\right \}.
\end{eqnarray}
We consider first the case of a bosonic system, in which case the
annihilation operators $\hat{\psi}_{{ i}}$, where the indices ${
i},{ j}={ a}, { b}, { b_2}, { ab}, { t}$ stand for atoms (A and
B), dimers (B$_2$ and AB) and trimers, satisfy standard bosonic
commutation relations. The terms proportional to $\chi_{i,j}'$
describe $s$-wave collisions between these species. We remark that
trimer formation via an atom-dimer resonance is actively studied
in ongoing experiments, and that the Feshbach-resonance-aided
photoassociation considered here could also be reformulated as a
laser frequency modulation scheme \cite{disp,Ling}. Finally, the
Bose-enhanced selectivity of dissociation channels in coherent
photodissociation of the heteronuclear trimers ABC was also
studied by Moore and Vardi \cite{PD}.

In the framework of a standard mean-field approach where
$\hat{\psi_{{ i}}}\rightarrow\sqrt{n}\psi_{{ i}}$ we find readily
\begin{eqnarray}
\dot{\psi}_{ a}&=&2i \sum_{ j} \chi_{{ a,j}} |\psi_{ j}|^2
\psi_{{ a}}+i\lambda\psi_{ b_2}^*\psi_{ t};\nonumber\\
\dot{\psi}_{ b}&=&2i \sum_{ j} \chi_{{ b,j}} |\psi_{ j}|^2 \psi_{{
b}}-i\Omega\psi_{ ab}^{*}\psi_{
t};\nonumber \\
\dot{\psi}_{ b_2}&=&2i \sum_{ j} \chi_{{ b_2,j}} |\psi_{ j}|^2
\psi_{ b_2}+i\lambda\psi_{ a}^*\psi_{ t};\nonumber\\
\dot{\psi}_{ ab}&=&2i \sum_{ j} \chi_{ ab,j} |\psi_{ j}|^2 \psi_{
ab}-i\Omega\psi_{ b}^*\psi_{ t}
+i(\Delta+\delta)\psi_{ ab},\nonumber\\
\dot{\psi}_{ t}&=&2i \sum_{ j} \chi_{{ t,j}} |\psi_{ j}|^2 \psi_{
t}+(i\delta\
-\gamma) \psi_{ t}+i\lambda\psi_{ a}\psi_{ b_2}\nonumber\\
&&-i\Omega\psi_{ b}\psi_{ ab},\nonumber\\
\end{eqnarray}
with $\chi_{ i,j}=n\chi_{ i,j}',~~\lambda=\lambda'\sqrt{n}$ and
$\Omega=\Omega'\sqrt{n}$, and the phenomenological decay rate
$\gamma$ accounts for the loss of intermediate trimers. The
maintenance of dark or unpopulated trimer state is important for
an efficient conversion even with a very short molecular lifetime
\cite{Ling}. For $\psi_b(0)=\psi_{ab}(0)=0$, Eqs.~(2) imply that
there is {\em no growth} in the populations of the atoms B and of
the dimers AB. This indicates that the mean-field Gross-Pitaevskii
equations, which are adequate to describe the short-time behavior
of the familiar coherent atom-dimer or atom-trimer conversion
\cite{Heinzen,Jing}, break down completely in studying the onset
of this bimolecular reaction.

A similar situation has been previously encountered in a broad
range of systems in quantum optics, but also in coupled degenerate
atomic and molecular systems such as the matter-wave superradiance
in Bose-condensed atoms \cite{QO,Uys,amplifier}. Following a
strategy developed in the study of these systems, we decompose the
problem into an initial stage dominated by quantum noise followed
by a classical stage that arises once the product components have
acquired a macroscopic population. The initial quantum stage is
treated in a linearized approach whose main purpose is to
establish the statistical properties of the initial fields
required for the classical stage \cite{amplifier}.

To simplify the description of the initial stages we note that in
the collisionless limit an effective second-quantized Hamiltonian
can be obtained by adiabatically eliminating the intermediate
excited state,
\begin{equation}
\hat{\cal H}_{\rm eff}=-(G \hat{c}_{ab}^\dag \hat{c}_b^\dag
\hat{c}_a \hat{c}_{b_2}+{h.c.})+\hat{c}_0,
\end{equation}
where $\hat{c}_0=\omega_1\hat{c}_a^\dag \hat{c}_{ a} \hat{c}_{
b_2}^\dag \hat{c}_{ b_2} +\omega_2 \hat{c}_{ab}^\dag \hat{c}_{ ab}
\hat{c}_{ b}^\dag \hat{c}_{ b},$
    \begin{eqnarray}
    G&=&(\lambda'\Omega'/\delta)\int dr \phi_{ ab}^*(r) \phi_{
    b}^* (r)\phi_{ a}(r) \phi_{ b_2}(r),\nonumber \\
    \omega_1&=&(\lambda'^2/\delta)\int dr \phi_{ a}^* (r)\phi_{
    a} (r)\phi_{ b_2}^* (r)\phi_{ b_2}(r),\nonumber \\
    \omega_2&=&(\Omega'^2/\delta)\int dr \phi_{ ab}^* (r)\phi_{
    ab}(r) \phi_{ b}^* (r)\phi_{ b}(r),\nonumber
    \end{eqnarray}
and $\hat{\psi}_{ i}(r,t)=\phi_{ i}(r)\hat{c}_{ i}(t)$.
Equation~(3) is reminiscent of the Hamiltonian describing
spin-exchange scattering in a two-species two-pseudospin-state
Bose condensate \cite{Wu}.

For short enough interaction times, the populations of the
products remain small compared to the total particle numbers. In
this regime, we can treat the fields ${\hat \psi}_{ a}$ and ${\hat
\psi}_{ b_2}$ classically, $\hat{c}_{ a, b_2}\rightarrow \sqrt{N_{
a,b_2}}$, and then neglect the term in Eq.~(3) describing only the
interactions of the modes $\hat{c}_{ a, b_2}$. This amounts to
linearizing the dynamics of the fields ${\hat c}_{ab}$ and ${\hat
c}_b$, with the noise source ${\hat f}_j^\dagger(t)$,
\begin{equation}\label{4}
\dot{{\hat c}}_{ab,b}(t)={\hat f}^\dag_{b,ab}(t)=i{\cal G}{\hat
c}_{b,ab}^\dag(t),
\end{equation}
which is familiar from quantum treatments of the optical
parametric oscillator \cite{QO} and of the molecular dissociation
(pair production) \cite{twin}. The noise operators satisfy
\begin{equation}\label{5}
\langle {\hat f}_{ i}^\dagger(t){\hat f}_{ j}(t') \rangle=0,~~
\langle {\hat f}_{ i}(t){\hat f}_{ j}^\dagger(t')\rangle ={\cal
G}^2\delta_{ij}\delta(t-t').\nonumber
\end{equation}
where ${\cal G}=G\sqrt{N_{ a}N_{ b_2}}$ and $i,j=ab$ or $b$ here
and in the following. It is these noise operators that trigger the
evolution of the system from initial vacuum fluctuations.

The quantum noise-induced populations of the $\hat{c}_{\rm g_2,b}$
mode and their fluctuations correlation are
\begin{eqnarray}
    N_j&\equiv& \langle \hat{c}_j^\dagger\hat{c}_j \rangle
    =\sinh^2({\cal G}t) \approx N_aN_{b_2}G^2t^2; \nonumber\\
    {\cal C}_{ab}={\cal C}_b &\equiv& \frac{\langle \Delta
    \hat{N}_{ab}\Delta \hat{N}_b \rangle}{\sqrt{N_{ab}N_b}}=1+\sinh^2({\cal G}t)>1,
\end{eqnarray}
where $\Delta \hat{N}_j\equiv\hat{N}_j-\langle \hat{N}_j \rangle$.
Equation~(5) can also be derived by solving Eq.~(3) to second
order in time. The second factorial moment of the modes
$\hat{c}_j$ is typical of chaotic fields, $g^{(2)}_j=2$, but they
are entangled, with
\begin{equation}
    \left [g^{(2)}_{ab,b}\right ]^2-g^{(2)}_{ab}g^{(2)}_b=
    \sinh^{-2}({\cal G}t)+4\sinh^{-1}({\cal G}t)>0,
\end{equation}
indicative of a violation of the classical Cauchy-Schwartz
inequality.

For comparison we comment briefly on the case where atoms A are
bosonic and atoms B fermionic. We obtain similar equations of
motion in that case, except that the noise operators
$(-\hat{f}_b^\dagger, \hat{f}_{ab}^\dagger)$ are now different.
These equations can be solved via a Bogoliubov transformation, and
we find that as a result of the Fermi statistics the
vacuum-noise-triggered population is now $N_j=\sin^2({\cal G}t)
<1$ (for the zero-momentum mode \cite{twin}), with the dimer-atom
pairs correlation becoming ${\cal C}_{ab}={\cal
C}_b=1-\sin^2({\cal G}t)<1$. The Mandel $Q$ parameters \cite{QO}
are
\begin{equation}
Q_j=\frac{\langle \hat{N}^2_j \rangle-{N}_j^2}{N_j} =\left \{
\begin{array}{l}
(a)~\cosh^2({\cal G}t)>1, \\
(b)~\cos^2({\cal G}t)<1,
\end{array}
\right.
\end{equation}
where (a) is for creating bosonic and (b) fermionic matter-wave
fields, which exhibit therefore super-Poisson or sub-Poisson
statistics \cite{QO}, respectively.

We conclude the discussion of the short-time dynamics by
mentioning that the long-time quantum statistics of the AB and B
populations, which are significantly influenced by the initial
vacuum fluctuations, can be calculated by a positive-$P$
representation technique \cite{Hope} and other methods \cite{Wu}.
Rather than adopting such a full quantum treatment, we proceed in
the following by combining the mean-field description of Eqs.~(2)
with stochastic classical seeds with statistics consistent with
the results of the short-time linearized quantum theory
\cite{amplifier}.

We now turn to the long-time reaction dynamics. We proceed by
numerically computing a large number of trajectories (typically
about 300) from initial classical seeds that satisfy the
short-time statistics of Eq.~(5). For each trajectory $n$, we use
Eqs.~(2) to calculate the particles populations $N_{i,n}(t)$ where
$i=$ A, B$_2$, AB, and B.

Figure~2 shows the standard deviation $\Delta N_i(t)$ of the
particle populations $i$=A, B, AB and B$_2$ for $\delta = \pm 3$,
and the insert shows a range $\pm \Delta N_i$ about their mean,
$\bar{N}_i(t)\pm \Delta N_i(t)$ for $\delta=3$ and in the case of
bosonic atoms. As expected, the small product seeds triggered by
the initial quantum fluctuations are significantly amplified
before reaching a stationary value for $\delta=3$.

\begin{figure}[ht]
\includegraphics[width=8cm]{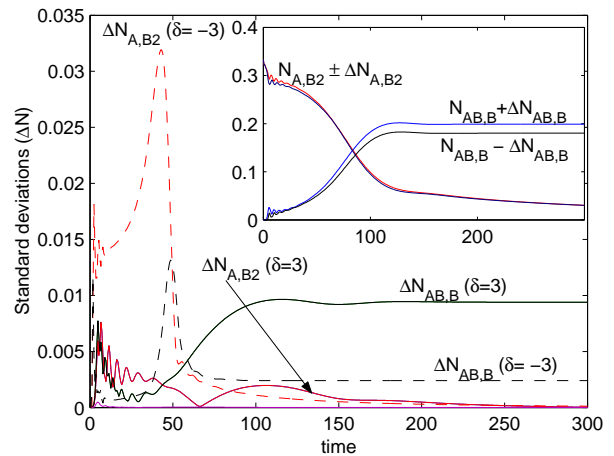} \caption{(Color online)
Standard deviations of the atom and dimer populations ($\Delta
N_j(t)={\langle \Delta \hat{N}_j^2\rangle}^{1/2}$) for $\delta=3$
or $\delta =-3$. Time is in units of $\lambda^{-1}$, and $\gamma
=1$. The other parameters are given in the text. The trimer number
remains zero at all times due to the CPT condition. Inset:
fluctuating range of the populations $\bar{N}_i(t)\pm \Delta
N_i(t)$ for $\delta=3$.}
\end{figure}

An important feature of the coherent abstraction reaction is that
it can be controlled and optimized by exploiting the existence of
a CPT dark state \cite{winkler,Jing}. Under a dynamical two-photon
resonance condition, Eqs.~(2) admit a steady-state CPT solution
such that a trimer state remains unpopulated at all times
\cite{later}
\begin{equation}
N_{ab,b}^s=\frac{2{\cal R}}{(1+{\cal R})\left [1+2{\cal
R}+\sqrt{(1-2{\cal R})^2+8{\cal R}\Omega^2/\lambda^2}\right ]},
\end{equation}
where we have applied the steady-state ansatz \cite{Jing} $
\psi_i^s=|\psi_i^s|e^{i\theta_i}e^{i\mu_it}$, $(\theta,\mu)_{b_2,
ab}=2(\theta,\mu)_b,(\theta,\mu)_a=(\theta,\mu)_b$ and ${\cal
R}\equiv {N_a(0)}/{2N_{b_2}(0)}$ is introduced to define the
initial ratio of the particles numbers. Using ${\partial
N_{ab}^s}/{\partial {\cal R}}=0$, we can find the maximum value
$N_{ab}^s|_{\rm max}= 1/3$ for ${\cal R}=1/2$.

Figure 3(a) shows the mean particle populations obtained by
averaging over $n_T= 300$ trajectories. In this specific example
atom A is $^{87}$Rb and atom B is $^{41}$K, $\lambda=
4.718\times10^4 $s$^{-1}$ and $ \Omega(t)=\Omega_{0}{\rm
sech}(t/\tau)$ with $\Omega_{0}/\lambda=20$, $\lambda\tau$=$20$
\cite{lens}. The collision parameters, in units of $\lambda/n$,
are $\chi_{aa}=0.5303$, $\chi_{bb} = 0.3214$, $\chi_{ab}=0.8731$,
and the others are taken as 0.0938 \cite{lens}. We note that the
scattering lengths of the various collisions, especially those
involving molecular trimers, are not known at this time. We have
therefore carried out numerical simulations by for a large set of
plausible collision parameters for the Rb-K, Rb-Na or other atomic
condensate \cite{later}. We found that the stable bimolecular
conversion is always possible for appropriate values of the
external field detuning $\delta$. The departure of the product
populations from the ideal CPT value is due to the fact that only
an approximate adiabatic condition exists for the CPT state
\cite{Ling,later}: $ \gamma_{\rm{nl}}(t)\approx
\frac{|\dot{\eta}|}{1+\eta}\frac{1}{4\lambda}\ll 1$, with
$\eta=\lambda/\Omega$, which becomes increasingly difficult to
satisfy in the last stages ($\eta\gg 1$).
\begin{figure}[ht]
\includegraphics[width=8cm]{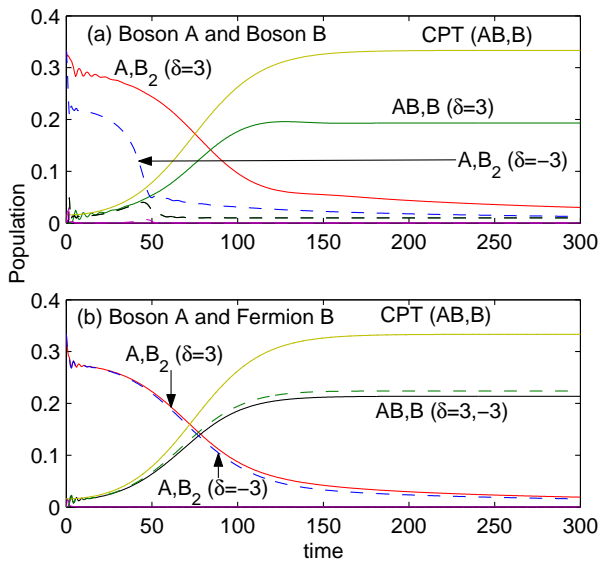} \caption{(Color online)
Populations of dimers and atoms for $\delta=3$ or $\delta =-3$ (in
units of $\lambda/n$) by averaging $300$ trajectories. Time is in
units of $\lambda^{-1}$, and $\gamma =1$. The trimer number
remains zero at all times. The line labelled "CPT" shows the ideal
population of products (dimers AB and atoms B). }
\end{figure}

Our proposal relies crucially on the capability to avoid rapid
collisional quenching or the formation of an unstable atom-dimer
sample. When energetically allowed, collision-induced reactions
always occur at some rate, and we need to guarantee that the time
scale over which quantum fluctuations dominate the dynamics is
short enough, so that the dynamics of the system is not
collision-dominated. From the condition $|{\cal G}t|<1$ we
estimate the maximum permissible collision-induced rate to be of
the order of $10^{5}$ s$^{-1}$ for $|\delta|=3$ and
$\Omega_0=20\lambda$. A typical low-temperature inelastic
collisions rate coefficient is $~ 10^{-17}m^3/s$
\cite{Hope,quench}, corresponding to a rate of about
$10^{3}$s$^{-1}$ for a sample density of $10^{14}/cm^3$. In that
case the fluctuations-induced dynamics will dominate the
short-time behavior of the system.

Finally we remark that by using $^{87}$Rb for the bosonic atoms A
and $^{40}$K for the fermionic atoms B we can realize the
conversion of bosonic pairs to fermionic pairs, which is likewise
described by Eqs.~(2) with the substitution
\begin{equation}\label{9}
\chi_{j,j}|\psi_{j}|^{2} \rightarrow A_j|\psi_{j}|^{4/3},
\end{equation}
where $A_j=(6\pi^2)^{2/3}/4M_j$, $j=ab, b$, and $M_j$ denotes the
particle mass, provided that we ignore the $s$-wave collisions of
fermionic particles and only consider their kinetic energy
\cite{LuLi}. In contrast to the purely bosonic case, molecule
formation is now found to be stable for both positive and negative
detunings, see Fig.~3(b).

In conclusion, we have shown that in the collective abstraction
reaction A+B$_2\rightarrow$ AB + B, the initial quantum
fluctuations lead to the strongly correlated creation of
dimer-atom pairs. This novel feature indicates that in the quantum
degenerate regime, elementary bimolecular reactive scattering are
significantly different from the familiar single-molecular
reaction, and may lead to fascinating new opportunities in
degenerate chemistry, such as the possible collective reaction
2A$_2 \rightarrow$ A$_3$ + A.

Future work will also study the unique "superchemistry" effects of
ultra-selectivity or confinement-induced stability in our system
\cite{PD}. In addition, a complete analysis of collisional effects
\cite{3body,Ni} will need to be considered. While experiments
along the lines of this analysis promise to be challenging, recent
progress in quantum degenerate chemistry \cite{winkler,PD,Ling}
and in the control of atom-molecule systems
\cite{trimer,tetramer,triple,08} indicates that achieving this
goal should become possible in the not too distant future.

\acknowledgements

This research is supported by the US Office of Naval Research, by
the US National Science Foundation, by the US Army Research
Office, and by the NSFC (10774047).


\begin{thebibliography}{99}
\bibitem{later} Details of the derivations will be
published elsewhere.
\end{thebibliography}
\end{document}